\documentclass[twocolumn]{autart}

\usepackage{graphicx}
\usepackage[cmex10]{amsmath}
\usepackage{amssymb}
\usepackage{mathtools}
\usepackage{csquotes}
\usepackage{verbatim}
\usepackage{mathrsfs}

\begin{document}

\begin{frontmatter}

\title{Regularized Nonparametric Volterra Kernel Estimation \thanksref{footnoteinfo}}

\thanks[footnoteinfo]{This paper was not presented at any IFAC 
meeting. Corresponding author G.~Birpoutsoukis. Tel. +32 (0)2 629 29 46}

\author[Brussels]{Georgios Birpoutsoukis}\ead{georgios.birpoutsoukis@vub.ac.be},
\author[Brussels]{Anna Marconato}\ead{Anna.Marconato@vub.ac.be},
\author[Brussels]{John Lataire}\ead{jlataire@vub.ac.be}, 
\author[Brussels]{Johan Schoukens}\ead{Johan.Schoukens@vub.ac.be}

\address[Brussels]{Department of ELEC, Vrije Universiteit Brussel, Pleinlaan 2, 1050 Brussel. \\ } 

\textit{This is a postprint copy of: Birpoutsoukis, G., Marconato, A., Lataire, J., \& Schoukens, J. (2017). Regularized nonparametric Volterra kernel estimation. Automatica, 82, 324-327. DOI: 10.1016/j.automatica.2017.04.014}


\begin{abstract}

In this paper, the regularization approach introduced recently for nonparametric estimation of linear systems is extended to the estimation of nonlinear systems modelled as Volterra series. The kernels of order higher than one, representing higher dimensional impulse responses in the series, are considered to be realizations of multidimensional Gaussian processes. Based on this, prior information about the structure of the Volterra kernel is introduced via an appropriate penalization term in the least squares cost function. It is shown that the proposed method is able to deliver accurate estimates of the Volterra kernels  even in the case of a small amount of data points.

\end{abstract}
 
\end{frontmatter}

\section{Introduction}
Modeling of nonlinear systems constitutes one of the most challenging topics in the field of system identification. In this work, nonparametric modeling is considered for a nonlinear system in the form of a time domain Volterra series \cite{schetzen1980}. The estimation of the Volterra series coefficients in an output error framework can be formulated as a Least Squares (LS) optimization problem. However, the number of coefficients to estimate is often too large for the LS problem to yield a solution of acceptable precision.

The problem of high variance has already been tackled successfully in \cite{chen2012},\cite{dinuzzo2015kernels},\cite{lataire2016},\cite{pillonetto2010},\cite{pillonetto2014} using regularization for the identification of a Finite Impulse Response (FIR) model for a Linear Time Invariant (LTI) system. In an output error framework, the linear-in-the-parameters FIR model can also be estimated by solving a linear LS problem. In order to reduce the variance of the estimated parameters, prior knowledge about the linear impulse response (exponential decay and smoothness) has been added to the parameter estimation problem. Moreover, Volterra kernel estimation using ridge regression as regularization method can be found in \cite{franz2006unifying}.

In this paper, the regularization method that has been proposed for estimating LTI FIR models is extended to the case of Nonlinear Time Invariant (NLTI) systems. The time domain truncated Volterra series is considered as the model for the underlying nonlinear process. Similarly to the linear case, it is assumed that the higher order kernels in the series are smooth and exponentially decaying. The main contribution of this work is to show that, by exploiting properties of the Volterra kernels in the higher dimensions, it is possible to construct a penalizing matrix into which the two aforementioned properties are encoded. A number of popular nonlinear model structures can be modeled with the proposed method, among which Wiener, Hammerstein and Wiener-Hammerstein.

In Section \ref{sec:Prelim} the problem formulation is given. In Section \ref{sec:Regularization}, the regularized LS estimators for FIR modeling are extended for the estimation of Volterra kernels. In Section \ref{sec:simulation}, the proposed approach is illustrated on a simulation example. Finally, conclusions are drawn in Section \ref{sec:Conc}.

\section{Problem formulation}
\label{sec:Prelim}
The considered nonlinear systems are modelled with the following model structure:

\begin{defn}
(Model Class) The input-output behaviour of the model satisfies the finite discrete-time Volterra series:

\begin{equation}
\begin{aligned}
y(k) &= h_{0} + \sum\limits_{m=1}^M H_{m} [ u ]_{k} + e(k) \\
H_{m} [u]_{k} &= \sum\limits_{\tau_{1}=0}^{n_{m}-1} \cdots \sum\limits_{\tau_{m}=0}^{n_{m}-1} h_{m}(\tau_{1}, \ldots, \tau_{m}) \prod\limits_{\tau = \tau_{1}}^{\tau_{m}} u(k - \tau)
\label{eq:General_Volterra}
\end{aligned}
\end{equation}

where $y(k)$ is the measured output at time instant k, $u(k)$ is the input signal applied to the system and the measurement noise $e(k)$ is a zero mean normally distributed i.i.d. signal, with variance $\sigma^{2}$, i.e. $e(k) \sim \mathcal{N} (0,\sigma^{2})$. $H_{m}[ \cdot ]$ is the m-th order Volterra operator and $n_{m} - 1$ corresponds to the number of past input values (the memory of $h_{m}$) needed to compute the system output. The constant term $h_{0}$ in \eqref{eq:General_Volterra} represents the zero-th order Volterra kernel.
\label{defn:1}
\end{defn}

\begin{rem}
In order to ensure the uniqueness of the identified Volterra kernels of order higher than one, the symmetrized kernels are considered \cite{schetzen1980}. For the second order kernel it means that $h^{0}_{2}(\tau_{1},\tau_{2}) = h^{0}_{2}(\tau_{2},\tau_{1}), \forall \tau_{1}, \tau_{2}$. It can be easily shown that the number of coefficients to be identified for a symmetric Volterra kernel of order $m \ge 1$, truncated at lag $n_{m}$, is given by $n_{\theta_{m}} = (1/m!) \prod\limits_{i = 0}^{m-1}(n_{m} + i)$. This is a straightforward result after considering all the possible permutations among the lag coordinates in the different dimensions.
\label{rem:unique}
\end{rem}

The problem is formulated as follows. Given $N$ measurements of $u(k)$ and $y(k)$, obtain an estimate of the Volterra kernels $h_{m}$. For convenience, \eqref{eq:General_Volterra} can be rewritten in vector notation as:

\begin{equation}
Y_{N} = \Phi_{N}^{T}\theta + E
\label{eq:LS}
\end{equation}

with (for $M = 2$ and $n_{1} = n_{2} = n$, i.e. $n_{\theta_{1}} = n_{1} = n$ and $n_{\theta_{2}} = \frac{n_{2}(n_{2}+1)}{2} = \frac{n^{2} + n}{2}$)

\begin{equation}\begin{aligned}
\Phi_{N} &= [\Phi^{T}_{0N} \  \Phi^{T}_{1N} \  \Phi^{T}_{2N}]^{T}, \ \ \theta = [\bar{\theta}_{0}^{T} \ \bar{\theta}_{1}^{T} \ \bar{\theta}_{2}^{T}]^{T} \\  
\bar{\theta}_{1} &= [h_{1}(0) \ h_{1}(1) \ \ldots \ h_{1}(n-1)]^{T}, \ \ \bar{\theta}_{0} = h_{0} \\ 
Y_{N} &= [y(n-1) \ y(n) \ \ldots \ y(N-1)]^{T} \\
\phi_{1}(k) &= [u(k) \ u(k-1) \ \ldots \ u(k-n+1)]^{T} \\
\Phi_{1N} &= [\phi_{1}(n-1) \ \phi_{1}(n) \ \ldots \ \phi_{1}(N-1)] \\
\Phi_{0N} &= [1 \ 1 \ 1 \ \ldots \ 1] \\
E &= [e(n-1) \ e(n) \ \ldots \ e(N-1)]^{T}
\end{aligned}
\end{equation}

where $\Phi_{0N} \in \mathbb R^{1 \times (N-n+1)}$, $\Phi_{1N} \in \mathbb R^{n_{\theta_{1}} \times (N-n+1)}$, $\Phi_{2N} \in \mathbb R^{n_{\theta_{2}} \times (N-n+1)}$, $\bar{\theta}_{0} \in \mathbb R$, $\bar{\theta}_{1} \in \mathbb R^{n_{\theta_{1}} \times 1}$, $\bar{\theta}_{2} \in \mathbb R^{n_{\theta_{2}} \times 1}$, $Y_{N} \in \mathbb R^{(N-n+1) \times 1}$, $\phi_{1}(k) \in \mathbb R^{n_{\theta_{1}} \times 1}$ and $E \in \mathbb R^{(N-n+1) \times 1}$. The matrix $\Phi_{2N}$ is constructed according to the vector $\bar{\theta}_{2}$ which concatenates $h_{2}(\tau_{1},\tau_{2})$ for $\tau_{1},\tau_{2} = 0, \ldots ,n_{2} - 1$. In this work, $\bar{\theta}_{2}$ is defined as $\bar{\theta}_{2} = [\bar{\theta}_{2,1} \ \bar{\theta}_{2,2} \ \cdots \ \bar{\theta}_{2,n_{\theta_{2}}}]^{T}$, where $\bar{\theta}_{2,i}, i=1,\ldots,n_{\theta_{2}}$ are arranged as depicted in Fig.~\ref{fig:Rotated_system}.

\begin{rem}
The vector $Y_{N}$ starts at $n-1$ and not $0$. The reason is that $n$ initial values of the input in $\Phi_{N}$ are not known.
\end{rem}

The total number of Volterra coefficients to be estimated is given by $n_{\theta} = 1 + n_{\theta_{1}} + n_{\theta_{2}}$.

\section{Regularized Volterra series estimation}
\label{sec:Regularization}

Given \eqref{eq:LS}, the regularized LS optimization problem with a quadratic penalty on the parameter vector $\theta$ is defined as:

\begin{equation}
\begin{aligned}
\hat{\theta}_{N}^{\textrm{Reg}} &= \arg\min_{\theta} \|Y_{N} - \Phi^{T}_{N}\theta\|^{2} + \theta^{T} D \theta \\
&= (\Phi_{N}\Phi^{T}_{N} + D)^{-1} \Phi_{N} Y_{N}
\end{aligned}
\label{eq:Costt_reg}
\end{equation}

with $D \in \mathbb R^{n_{\theta} \times n_{\theta}}$. For $D = 0$ the LS cost function is obtained which corresponds to the Maximum Likelihood (ML) estimation of the parameter vector $\theta$ \cite{ljung1999system},\cite{pintelon2012system}. Typically, when the Volterra series is used to model nonlinear dynamics, either the number of coefficients to be estimated exceeds the number of available measurements ($n_{\theta} > N$) or the amount of measured data $N$ is not sufficient to obtain an estimated model of desired accuracy. In the first case when $n_{\theta} > N$ the LS solution is not even unique and the ML approach cannot be used. In this work these problems are tackled with the use of regularization ($D \neq 0$). 

By a proper structuring of $D$ in \eqref{eq:Costt_reg}, one can impose prior knowledge about the true system. In the linear case the penalty $D$ can be tuned using a Bayesian perspective and considering the impulse response as a realization of a random Gaussian process \cite{rasmussen2006},\cite{pillonetto2010},\cite{pillonetto2011}. Assuming $\theta \sim \mathcal{N} (0,P)$ and taking \eqref{eq:LS} into account, then $\theta$ and $Y_{N}$ are jointly distributed Gaussian variables and $\theta \bigl| Y_{N} \sim \mathcal{N} (\hat{\theta}_{N}^{\textrm{apost}},P^{\textrm{apost}})$ with:

\begin{equation*}
\hat{\theta}_{N}^{\textrm{apost}} = (\Phi_{N}\Phi^{T}_{N} + \sigma^{2} P^{-1})^{-1} \Phi_{N} Y_{N}
\end{equation*}

where $\hat{\theta}_{N}^{\textrm{apost}}$ denotes the Maximum A Posteriori (MAP) estimate of $\theta$. It can be observed that $\hat{\theta}_{N}^{\textrm{apost}} = \hat{\theta}_{N}^{\textrm{Reg}}$ in \eqref{eq:Costt_reg} if $D = \sigma^{2} P^{-1}$. This perspective offers an alternative interpretation of the regularization matrix $D$ which can be constructed by tuning the prior covariance matrix $P$. Throughout the paper, the prior covariance matrix for the Volterra kernel of order $m$ is denoted by $P_{m}$.

\subsection{Regularization for the first order Volterra kernel}
\label{subsec:Reg_first}

For the first order kernel, prior information about smoothness and exponential decay of the impulse response can be imposed by the so-called Diagonal/Correlated (DC) \cite{chen2012},\cite{chen2016kernel},\cite{pillonetto2014} structure of $P_{1}$:

\begin{equation}
P_\textrm{1}(i,j) = c \cdot e^{-\alpha \lvert i - j \rvert} \ e^{-\beta \frac{ (i + j)}{2}}
\label{eq:PDC}
\end{equation}

where $P_{1}(i,j) = E[h_{1}(\tau_{i}) \ h_{1}(\tau_{j})]$, $h_{1}(\tau_{i})$ is the first order impulse response coefficient at lag $\tau_{i}$, $E[ \cdot ]$ denotes the expected value and $0 \le \alpha, \beta \le \infty$. The parameters $c$, $\alpha$ and $\beta$ are called hyper-parameters and they are computed by maximizing the marginal likelihood of the observed output \cite{pillonetto2014},\cite{rasmussen2006},\cite{pillonetto2015tuning}.

\subsection{Extension to the second order Volterra kernel}
\label{sec:Volterra_reg}

\begin{assum}
The second order Volterra kernel is smooth and decaying. Moreover, no prior knowledge on the correlation between Volterra coefficients of different order is available.
\end{assum}

In Fig.~\ref{fig:Smooth_decaying_kernel}, an example of a decaying and smooth second order Volterra kernel has been constructed. It is the kernel of a nonlinear Wiener system where a stable linear system is cascaded by a quadratic static nonlinear system \cite{schetzen1980}.

Given the cost function \eqref{eq:Costt_reg} and $\theta = [\bar{\theta}_{0}^{T} \ \bar{\theta}_{1}^{T} \ \bar{\theta}_{2}^{T}]^{T}$, $D$ is a block-diagonal matrix with elements $D_{0}$, $D_{1}$ and $D_{2}$ on the diagonal. $D_{0}$ is a scalar defined by $D_{0} = \sigma^{2}P_{0}^{-1}$ where $P_{0}$ represents the variance of the Gaussian random variable $h_{0}$. The penalizing matrix for the first order kernel is $D_{1} = \sigma^{2}P_{1}^{-1}$ and $P_{1} \in \mathbb R^{n_{\theta_{1}} \times n_{\theta_{1}}}$ is defined as described in Section \ref{subsec:Reg_first}. Finally, the covariance matrix $P_{2} \in \mathbb R^{n_{\theta_{2}} \times n_{\theta_{2}}}$ such that $D_{2} = \sigma^{2}P_{2}^{-1}$ is developed in this paper.

Given a second order Volterra kernel, the covariance matrix $P_{2}$ should satisfy the following:

\begin{figure}
	\centering
{\includegraphics[scale=0.23]{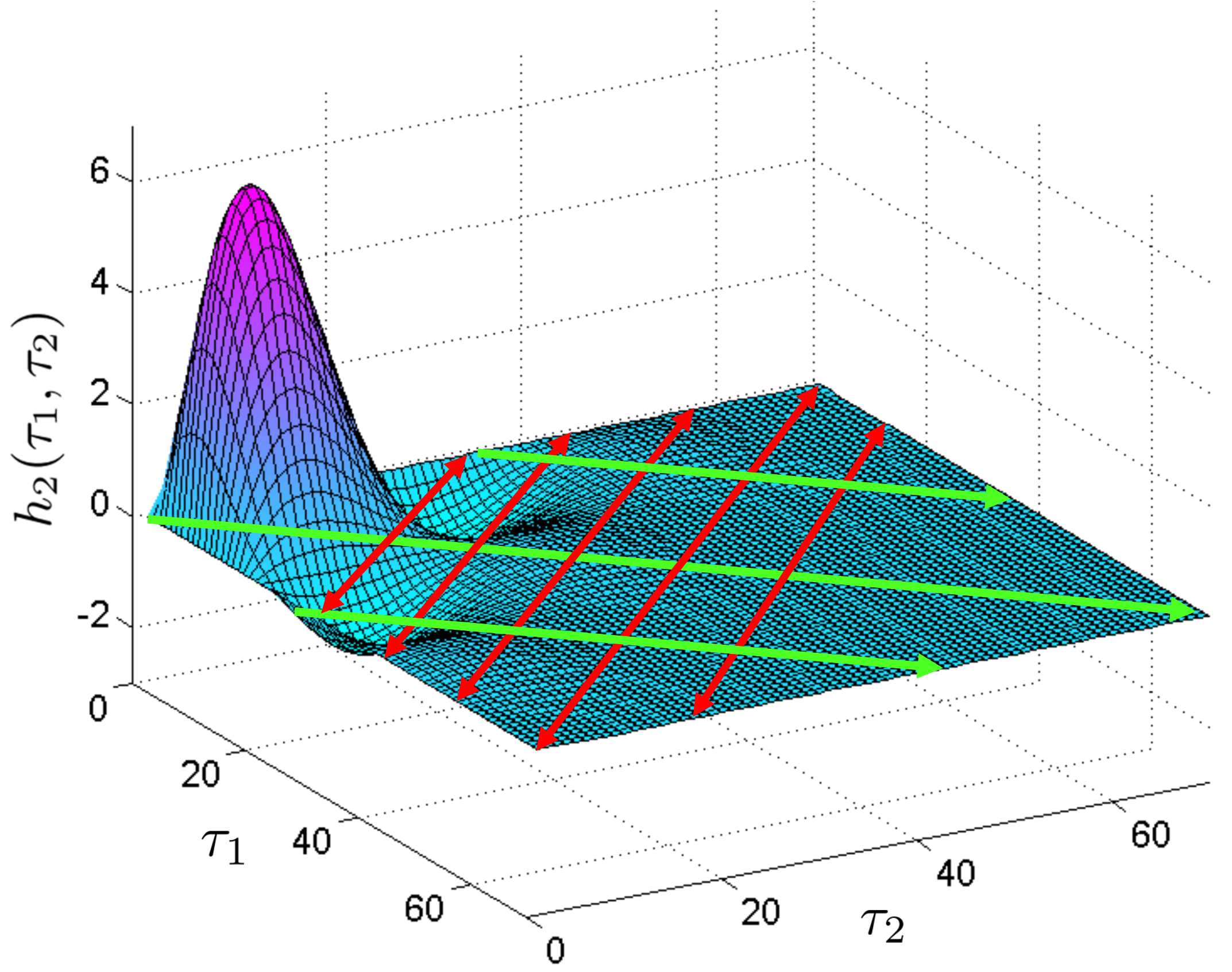}}
	\caption{An exponentially decaying and smooth second order kernel together with the two perpendicular regularizing directions (green and red).}
	\label{fig:Smooth_decaying_kernel}
\end{figure}

\begin{property}
It should describe the fact that the Volterra kernel decays along any possible direction and, moreover, neighbouring coefficients of the Volterra kernel are correlated.
\label{prop1}
\end{property}
\begin{property}
The considered second order Volterra kernel is symmetric which implies that $h_{2}(\tau_{1},\tau_{2}) = h_{2}(\tau_{2},\tau_{1}), \forall \tau_{1}, \tau_{2}$. Therefore $P_{2}$ should remain unaltered after interchanging the coordinates $\tau_{1}$ and $\tau_{2}$.
\label{prop2}
\end{property}
\begin{property}
Finally, as for any covariance matrix, $P_{2}$ should be constructed to be a symmetric positive-semidefinite matrix.
\label{prop3}
\end{property}

These three properties reduce the freedom to construct an appropriate covariance matrix for the Volterra kernel.

\subsubsection{The two directions and the covariance matrix}
\begin{figure}
	\centering
{\includegraphics[scale=0.2]{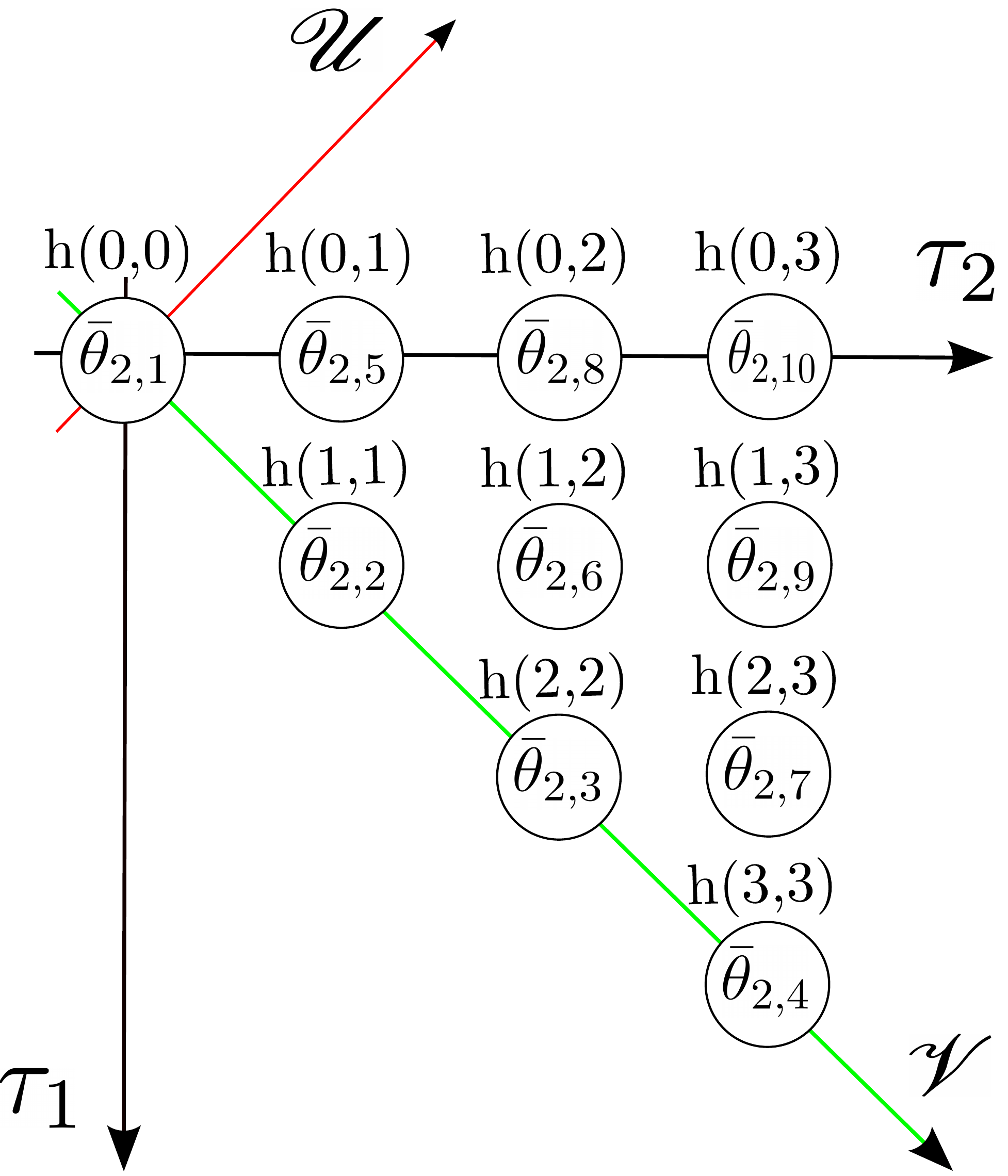}}
	\caption{The initial coordinate system ($\tau_{1}$,$\tau_{2}$) together with the rotated one ($\mathscr{V}$,$\mathscr{U}$), used in the covariance matrix of the second order kernel.}
	\label{fig:Rotated_system}
\end{figure}

Two directions in the $\tau_{1} - \tau_{2}$ plane are chosen in order to build a proper covariance matrix for the second order kernel. The first direction, along which prior information about the kernel will be imposed, is chosen to be the $\mathscr{V}$ diagonal. It is depicted in Fig.~\ref{fig:Smooth_decaying_kernel} and Fig.~\ref{fig:Rotated_system} as the green diagonal direction. The second direction is the $\mathscr{U}$ anti-diagonal direction and it is represented by the red lines in Fig.~\ref{fig:Smooth_decaying_kernel} and Fig.~\ref{fig:Rotated_system}. In this way, it is possible to construct a covariance matrix that will satisfy Property \ref{prop1}.

To this aim, a new rotated coordinate system is defined as depicted in Fig.~\ref{fig:Rotated_system}. Using the new system ($\mathscr{V}$,$\mathscr{U}$) it becomes more clear how the properties of exponential decaying and smoothness can be described in these directions, and further how to extend the DC covariance matrix structure \eqref{eq:PDC} to dimensions higher than one. 

\begin{defn}
Given the elements of a symmetric second order kernel $h_{2}(\tau_{1},\tau_{2})$, vectorized in $\bar{\theta}_{2}$, and the coordinate system ($\mathscr{V}$,$\mathscr{U}$) depicted in Fig.~\ref{fig:Rotated_system}, the covariance matrix $P_{2} = E[\bar{\theta}_{2} \bar{\theta}_{2}^{T}]$ used to describe and further penalize the coefficients of the second order Volterra kernel is given by:

\begin{equation}
\begin{aligned}
P_{2}(i,j) &= c_{2} \ p_{\mathscr{V}}(i,j) \ p_{\mathscr{U}}(i,j) \\
p_{\mathscr{V}}(i,j) &= e^{-\alpha_{\mathscr{V}} \bigl| \lvert \mathscr{V}_{i} \rvert - \lvert \mathscr{V}_{j} \rvert \bigl|} \ e^{-\beta_{\mathscr{V}} \frac{ \bigl| \lvert \mathscr{V}_{i} \rvert + \lvert \mathscr{V}_{j} \rvert \bigl|}{2}} \\
p_{\mathscr{U}}(i,j) &= e^{-\alpha_{\mathscr{U}} \bigl| \lvert \mathscr{U}_{i} \rvert - \lvert \mathscr{U}_{j} \rvert \bigl|} \ e^{-\beta_{\mathscr{U}} \frac{ \bigl| \lvert \mathscr{U}_{i} \rvert + \lvert \mathscr{U}_{j} \rvert \bigl|}{2}}
\end{aligned}
\end{equation}

with $\mathscr{V}_{i}$ and $\mathscr{U}_{i}$ the coordinates of $\bar{\theta}_{2,i}$ on the $\mathscr{V}$- and $\mathscr{U}$-axes.

\end{defn}

The hyper-parameters $\alpha_{\mathscr{V}}$ and $\beta_{\mathscr{V}}$ are connected to the smoothness and the decay of the kernel in the diagonal direction, respectively. Similarly, $\alpha_{\mathscr{U}}$ and $\beta_{\mathscr{U}}$ are linked to the correlation between the Volterra coefficients as well as the decay in the $\mathscr{U}$ direction. The hyper-parameter $c_{2}$ represents a scaling factor. The matrix $P_{2}$ is a valid covariance matrix according to the results of \cite{christakos2000norm}, section 3, on valid covariance matrices with Non-Euclidean norms.

The second order Volterra kernel of the underlying true system is assumed to be a realization of a Gaussian process with zero mean and covariance matrix given by $P_{2}$. The hyper-parameters used to construct the matrix $P$, and further $D = \sigma^{2} P^{-1}$, are determined also in this case by maximizing the marginal likelihood for the observed output.

\begin{rem}
The proposed covariance matrix $P_{2}$ satisfies Property \ref{prop2}. It can be easily seen that interchanging the coordinates $\tau_{1}$ and $\tau_{2}$ corresponds to a change in the sign of the coordinate $\mathscr{U}$ in $p_{\mathscr{U}}$. Given that the absolute values of the coordinates are considered, the covariance matrix remains in this case unaltered, as required. Moreover, $P_{2}(i,j) = P_{2}(j,i)$ therefore the proposed matrix satisfies also Property \ref{prop3}.
\end{rem}

\section{Numerical example}
\label{sec:simulation}

\begin{figure}
	\centering
{\includegraphics[scale=0.32]{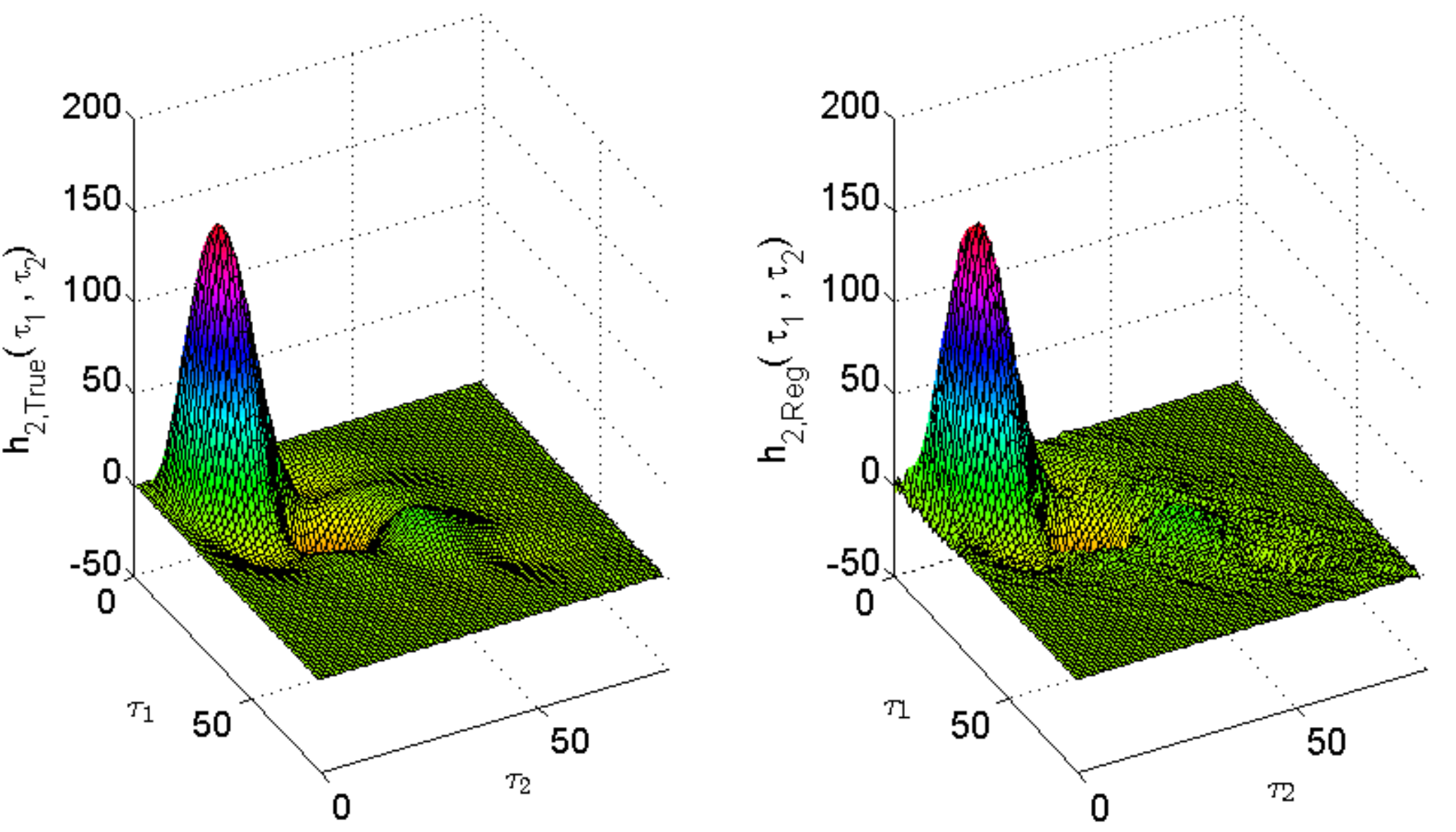}}
	\caption{Left: True second order Volterra kernel. Right: Estimated kernel for $\textrm{SNR} = 20 \textrm{dB}$ and $N/n_{\theta} = 1.3$.}
	\label{fig:Example}
\end{figure}

The method is illustrated on the system shown in Fig.~\ref{fig:Numer_ex} with:

\begin{figure}
	\centering
{\includegraphics[scale=0.28]{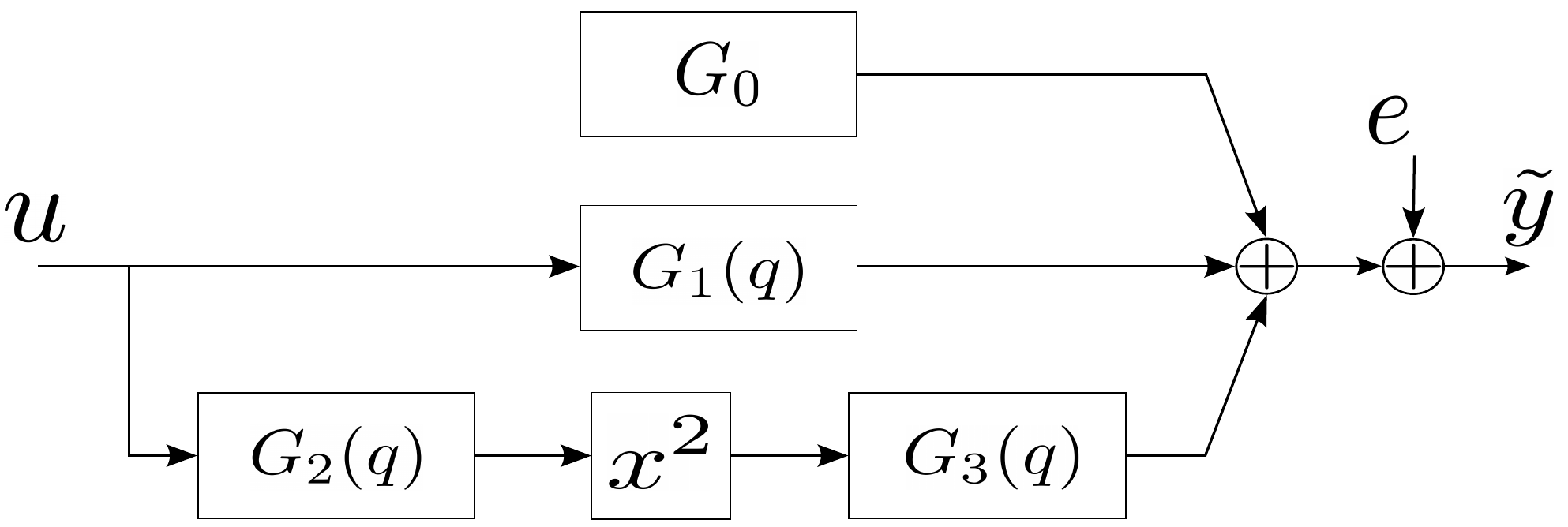}}
	\caption{The system chosen for the numerical illustration. A Wiener-Hammerstein nonlinear structure is considered.}
	\label{fig:Numer_ex}
\end{figure}

\begin{equation*}
\begin{aligned}
G_{0} &= 2, \ \ G_{1}(q) = \frac{0.7568 q^{-1}}{1 - 1.812 q^{-1} + 0.8578 q^{-2}} \\
G_{2}(q) &= \frac{1.063 q^{-1}}{1 - 1.706 q^{-1} + 0.7491 q^{-2}}, \ \ G_{3}(q) = 1.5 G_{1}(q)
\end{aligned}
\end{equation*}

where $q^{-1}$ is the delay operator ($q^{-1} u(k) = u(k-1)$). The input-output behavior of this system can be modelled with a Volterra series of degree two (eq.\eqref{eq:General_Volterra} with $M = 2$). The systems $G_{2}(q)$ and $G_{3}(q)$ have been chosen such that the second order Volterra kernel, associated with the system in Fig.~\ref{fig:Numer_ex}, has different properties of smoothness and decay in the directions $\mathscr{U}$ and $\mathscr{V}$ (Fig.~\ref{fig:Rotated_system}). Even though the zero-th, first and second order Volterra kernels are identified simultaneously, we emphasize here on the identification of the second order Volterra kernel. The second order kernel corresponding to the considered system is depicted in Fig.~\ref{fig:Example} (on the left) and has been constructed using the measurement method described in \cite{schetzen1980}, pages 44-47.

The input signal is a filtered random phase multisine of unit power, and the noise term $e$ is a zero mean normally distributed i.i.d. signal. The method is tested for a Signal-to-Noise Ratio (SNR) of $20$ dB \footnote{SNR is defined as $\frac{\textrm{var}(y^{0})}{\textrm{var}(e)}$ where $y^{0}$ denotes the noiseless system output and var is the variance}. The results are reported with respect to the ratio $N/n_{\theta}$. The estimated second order Volterra kernel for $N/n_{\theta} = 1.3$ and truncation at $n_{1} = n_{2} = 80$ lags is shown in Fig.~\ref{fig:Example}. 

The method is also illustrated for different values of the ratio $N/n_{\theta}$ (here $n_{1} = n_{2} = 80$ lags are fixed therefore $n_{\theta}$ is fixed and $N$ is increasing). For each value $N/n_{\theta}$, 100 Monte Carlo (MC) simulations are performed. In every MC simulation, a different excitation signal $u$ and noise realization $e$ are used. The ability of the estimated model to simulate the output of the system is measured with respect to a noiseless validation data set of $N_{\textrm{val}} = 50000$ data points, which also changes in every MC iteration. The portion of validation data not described by the modeled output is measured by $\textrm{Err}_\textrm{val} = \textrm{rms}(y_{\textrm{val}} - \hat{y}_{\textrm{val}})/\textrm{rms}(y_{\textrm{val}})$, where $y_{\textrm{val}} \in \mathbb R^{N_{\textrm{val}}}$ denotes the noiseless validation output, $\hat{y}_{\textrm{val}} \in \mathbb R^{N_{\textrm{val}}}$ is the output of the estimated model when the latter is excited by the validation input signal $u_{\textrm{val}} \in \mathbb R^{N_{\textrm{val}}}$ and $\textrm{rms}$ denotes the root mean square.

\begin{figure}
	\centering
{\includegraphics[scale=0.42]{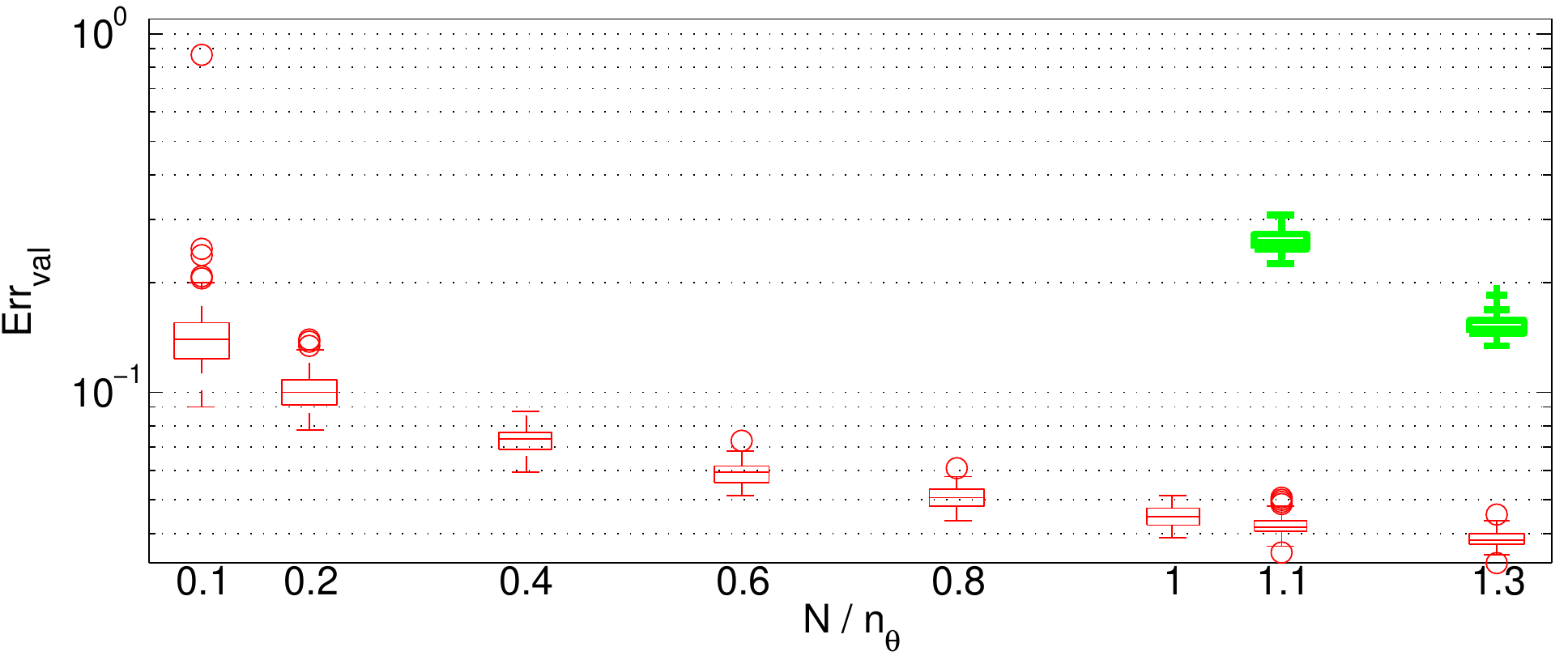}}
	\caption{Box-plot of the Monte Carlo simulation results for the considered nonlinear system. Black thin line boxes: Regularized estimation Grey wide line boxes: Unregularized estimation (LS solution).}
	\label{fig:Monte_Carlo}
\end{figure}

It is clear in Fig.~\ref{fig:Monte_Carlo} that the regularized Volterra kernel estimation (red boxes) produces models which describe more than 90\% of the noiseless validation output when $N/n_{\theta} > 0.2$, even though the noise level is quite high. Moreover, the error in the simulated output decreases with an increasing number of data used for estimation, as expected. The LS solution obtained for $N/n_{\theta} \geq 1$ (unregularized estimation) is also computed and the MC results (green boxes) are compared with the ones obtained with the proposed method. It is clear that the models obtained with the LS approach (ML estimation), given the high number of parameters (truncation at $n_{1} = n_{2} = 80$ lags corresponds to $3321$ Volterra coefficients) to be estimated and the limited data length (maximum $N/n_{\theta} = 1.3$ in Fig.~\ref{fig:Monte_Carlo}), are less able to simulate the system output than the ones computed when regularization is used. It is important to note that, qualitatively, the same conclusions can be derived if different values of SNR or another structure for the second order nonlinearity (e.g. Hammerstein case) are considered.

\section{Conclusions}
\label{sec:Conc}

The method presented in this paper constitutes a way to include prior information about the Volterra kernels (smoothness and decay) for the identification of nonlinear dynamics, by penalizing properly the higher dimensional impulse response coefficients. Even in the situation when a small amount of measurements is available (smaller than the number of parameters) and the LS approach cannot be used, modeling with the Volterra series is feasible. The regularization method presented in this paper renders the modeling of a big class of nonlinear systems with the Volterra series possible, which has not been the case before. However, several issues should be adressed in the future, such as efficient ways to optimize the hyperparameters of the regularization matrix in order to reduce the risk of resulting in a local minimum of the marginal likelihood.

\begin{ack}                               
This work was supported in part by the Fund for Scientific Research (FWO-Vlaanderen), by the Flemish Government (Methusalem), the Belgian Government through the Inter university Poles of Attraction (IAP VII) Program, and by the ERC advanced grant SNLSID, under contract 320378.
\end{ack}

\bibliographystyle{plain}
\bibliography{Automatica2017}   
\end{document}